\documentclass[twocolumn,pre,twoside,showpacs,preprintnumbers,floatfix]{revtex4}
\usepackage{graphicx}
\usepackage{dcolumn}
\usepackage{bm}
\begin{document}

\title{Effective Interaction of Charged Platelets in Aqueous Solution:
\\Investigations of Colloid Laponite Suspensions by Static Light Scattering 
and Small-Angle X-Ray Scattering }

\author{Li Li$^a$, L. Harnau$^b$, S. Rosenfeldt$^a$,  
and M. Ballauff$^a$}
\email[]{E-mail: Matthias.Ballauff@uni-bayreuth.de}
\affiliation{$^a$Physikalische Chemie I, University of Bayreuth, 
95440 Bayreuth, Germany\\
$^b$Max-Planck-Institut f\"ur Metallforschung,  
Heisenbergstr.\ 3, D-70569 Stuttgart, Germany, 
\\and Institut f\"ur Theoretische und Angewandte Physik, 
Universit\"at Stuttgart, Pfaffenwaldring 57, D-70569 Stuttgart, Germany}

\date{\today}

\begin{abstract}
We study dilute aqueous solutions of charged disk-like mineral particles (Laponite) by a combination of static light scattering (SLS) and small-angle 
x-ray scattering (SAXS). Laponite solutions are known to form gels above a certain 
critical concentration that must be described as non-equilibrium states. Here we 
focus on the investigation by SLS and SAXS at concentrations below gelation 
($c < 0.016$ g/L) and at low concentrations of added salt ($0.001$ and $0.005$ mM). Thus,  
we have obtained the scattering function of single Laponite platelets as well as the 
structure factor describing their interaction at finite concentration. A detailed 
analysis of the combined sets of data proves that the solutions are in a well-defined 
equilibrium state. Moreover, this analysis demonstrates the internal 
consistency and accuracy of the scattering functions obtained at finite concentrations.  
We find that Laponite particles interact through an effective pair potential that 
is attractive on short range but repulsive on longer range. This finding demonstrates 
that Laponite solutions exhibit only a limited stability at the concentration of 
added salt used herein. Raising the ionic strength to $0.005$ mM already leads to 
slow flocculation as is evidenced from the enhanced scattering intensity at smallest 
scattering angles. All data strongly suggest that the gelation occurring at higher 
concentration is related to aggregation. 
\end{abstract}

\maketitle

\section{Introduction} 
Laponite consists of disk-like clay particles with a thickness 
$L \approx 1$ nm and a radius  $R \approx 12.5$ nm. Each platelet carries a few 
hundred elementary charges. For a long time, aqueous suspensions of Laponite 
have been extensively investigated as a model system for disk-like colloids 
\cite{Ram:86,Ram:90,Ram:93,Mou:95}. A theoretical description of such a 
well-characterized model clay appeared feasible. Early attempts focussed on 
the structure of the electric double layer around a single platelet
\cite{chan:94,triz:97a,triz:97b}, or around two parallel platelets \cite{delv:99}.
Much of the recent work was initiated by the theoretical treatment by Hansen 
and coworkers \cite{dijk:97}. In the model of Hansen et al. Laponite platelets 
and their associated electric double layers are represented by nonintersecting 
discs carrying a constant electrostatic quadrupole moment. The spatial arrangement
of these objects has been investigated by Monte Carlo simulations. The results of 
these studies point to a reversible sol-gel transition at platelet number density 
$n>0.06\, R^{-3}$ (corresponding to a platelet concentration of 3.4 \% by weight), 
where the structure of the gel phase is strongly reminiscent of a "house-of-cards" 
\cite{Olp:77}. However, the quadrupolar disc model has turned out to be an 
oversimplification. In particular it is inadequate at very short range, where any 
multipolar expansion breaks down, and at long range, where electrostatic 
interactions between the platelets are exponentially screened. 

A screened electrostatic potential between two arbitrary oriented, charged 
platelets has been worked out within the framework of a linearized 
Poisson-Boltzmann theory \cite{hsu:97,rowa:00,triz:02,agra:04}. At a fixed 
center-to-center distance, this electrostatic potential (always repulsive) 
is maximized for co-planar platelets, which corresponds to the maximum overlap 
of electric double layers, and minimized when the platelets are co-axial and 
parallel. An intermediate potential is found for T-shape perpendicular platelets. 
At present only preliminary results of Monte Carlo simulations for platelets 
interacting through such an anisotropic potential are available, indicating 
that the platelet fluid is in the isotropic phase for platelet number density 
$n<0.1\, R^{-3}$ \cite{triz:02}.

Later an interaction site model of Laponite was 
put forward, and investigated by molecular-dynamics simulations \cite{kutt:00} and 
integral equations \cite{harn:01}. In this model each platelet carries discrete 
charged sites rigidly arranged on an array inscribed in the disc. Sites on different 
platelets interact via a repulsive screened Coulomb interaction. Upon varying the 
platelet number density and screening length, a sol, gel, and crystal phase have 
been identified. Additional rim charges of opposite sign have been found to lead to 
T-shaped pair configurations and clustering of the platelets. It should be noted, 
however, that these models treat the Laponite particles as uniform.

Summing up, several mechanisms have been proposed as origin for the gel formation 
in Laponite solutions by now: First of all, the classical answer to this problem is the 
formation of a "house of cards"-structure by attractive interactions between the 
platelets, possibly caused by positive charges on the rims \cite{Olp:77}. 
This includes more recent approaches taking into account the charge distribution 
around the platelets \cite{harn:01}.

As an alternative to this modeling, it was argued that this type of aggregation will 
only take place at higher ionic strength ($>0.01$ mM) and may be ruled out as cause 
for gelation. Instead, the repulsive interaction between the charged platelets leads 
to a glassy state 
\cite{Mou:95,Kro:96,Mou:98a,Mou:98b,Kro:98,Bon:99a,Bon:99b,Kna:00,Bon:01,Bon:02,Bell:03}. 
Slow time-dependent processes within the gel were interpreted in terms of an aging of 
the glass. Very recently, it has been hypothesized that Laponite suspensions behave 
like an attractive glass \cite{Tan:04,Tan:05}. A comprehensive discussion of the 
treatment of possible glassy states in Laponite suspensions including a comparison 
with experimental data was  recently given by Mongondry et al. \cite{Mon:05}.  

A markedly different model recently proposed by Nicolai et al. invokes aggregation 
in order to explain the process of gelation \cite{nico:01a,nico:01b,Mon:04}. Weak 
attraction may lead to a slow aggregation of the particles even at low ionic strength.
As a consequence of this, no well-defined phases or phase transitions can be identified 
anymore. The observations made by Nicolai et al. \cite{Mon:05} are in agreement with 
a study of Martin et al. \cite{Mar:02}. Hence, the origins of the sol-gel transition 
have not yet been fully clarified despite of more than two decades of research. 

Most of the theoretical expositions on gelation and structure formation in Laponite 
suspensions have assumed a system of monodisperse platelets, i.e., thin disks with a 
radius of $12.5$ nm and a thickness of $1$ nm. However, it has become possible to 
visualize directly the Laponite particles by various techniques as AFM or electron 
microscopy. From these micrographs it is obvious that Laponite exhibits a considerable 
polydispersity \cite{Bih:01,Bal:03,Her:04,leac:05}. A second point that is common to 
many experimental study is the analysis of obvious non-equilibrium states reached after 
passing the gel line. As the matter of fact, considerable efforts have been devoted 
to the investigation of the exact position of the gel line. The experimental 
difficulties of the investigations of such time-dependent states are at hand. Moreover, 
no clear information on the nature of the interparticle potential can be derived from 
essentially non-equilibrium states. An advanced understanding of Laponite suspensions, 
however, requires a clear delineation of the range of repulsion together with a clear
assessment of a possible attractive part of the potential that may come into play at 
short distances \cite{Olp:77}. 

In this paper we present a comprehensive study of Laponite suspensions in the dilute 
regime by static light scattering (SLS) and small-angle x-ray scattering (SAXS). We 
purposely avoid higher concentrations leading to gelled states. Instead, particular 
care is devoted to the proof that we do look at equilibrium states indeed. Hence, data 
are taken at different small concentrations of the solute and extrapolated to infinite
dilution. This leads to the scattering intensity $I_0(q)$ of single platelets as the 
function of $q$, the magnitude of the scattering vector
$\vec q$ ($q = (4 \pi/ \lambda) \sin(\theta / 2)$; $\lambda$: wavelength of radiation; 
$\theta$: scattering angle). $I_0(q)$ in turn can be modelled to give the polydispersity. 
The scattering intensity $I(q)$ obtained at finite but small concentrations is used 
to determine the "measured structure factor" $S_M(q)$ which is related to the pair 
correlation function $g(r)$ and the interaction potential $V(r)$. However, the 
appreciable polydispersity of the Laponite particles precludes a direct analysis of 
$S_M(q)$ to yield $V(r)$. Here we shall address this problem within the framework of 
the "Polymer reference interaction site model" (PRISM). Using the approach developed 
previously for suspensions of bidisperse platelets \cite{harn:02}, size polydispersity 
of the platelets will be taken into account by a multicomponent PRISM integral equation 
theory. We shall demonstrate that the combination of precise scattering data with the 
multicomponent PRISM theory leads to an unambiguous assessment of the interaction 
potential in solution.

The paper is organized as follows: After the section Experimental we shall delineate 
in section III the theory of light scattering of dilute suspension. In particular, 
the check of the measured data for consistency is given. This section contains also 
the description of the multicomponent interaction site model used for calculating
structure factors. Section IV will give describe the comparison with measured 
scattering data and the resulting effective interaction potential. A final brief 
section V will conclude this paper.

\section{Experimental}
\subsection{Materials}
Laponite RD  was obtained from Laporte Industries Ltd. and is used for 
our study without further purification. Laponite powder was added to a $10^{-3}$ mol/l 
NaCl solution in distilled water and mixed at high speed with a magnet stirrer for 4 
hours. The dispersions were initially turbid, but became clear in 15-20 minutes, 
indicating that the particles were fully hydrated. All Laponite samples (volume fraction 
ranging from \mbox{0.02 \%} to \mbox{0.16 \%}) were prepared separately, without using a 
stock solution. Vigorous stirring of the solutions was found to be necessary for dispersing particles homogeneously. SAXS-experiments demonstrated that incomplete dissolution leads 
to the formation of clusters and large aggregates. 

The pH values of all the samples were adjusted to 10 by using NaOH in order to avoid 
degradation or dissolution of Laponite particles. Dispersions were kept still overnight 
before any use in order to achieve equilibrated state. For SAXS experiments the samples 
were measured without further treatment, as no difference was observed between the 
scattering intensities of unfiltered dispersion and filtered one by using 0.22 mm 
filters (Millipore). For static light scattering experiments the dispersions were 
filtered by using 0.22 mm filters (Millipore) into quartz cuvettes to eliminate 
dust and remaining aggregates.

\subsection{Methods}
The scattering intensity $I(q)$  was obtained by static light scattering in the region 
of smallest $q$-values and by SAXS for higher $q$-values. Static light scattering were 
performed by a Sofica SLS-Spectrophotometer, using vertically polarized light with a 
wavelength of 632.8 nm. The sample is contained in a cylindrical 10 mm diameter quartz 
cuvette, which is immersed in an index-matching toluene bath of 80 mm diameter. Scattered 
light is detected by a photomultiplier, which is mounted on a goniometer arm. The 
angle range is 35-125 deg., corresponding to q range of 
$7.9 \times 10^{-3} - 2.3 \times 10^{-2}$ nm$^{-1}$.

For SAXS experiments, the data are composed by two sets in different q ranges. One set 
of data in relatively lower q range were measured at the European Synchrotron Radiation 
Facility (ESRF) in Grenoble on the high brilliance beam line ID2. The instrumental 
set-up for Laponite samples is as following: wavelength $\lambda$ = 0.0995 nm with 
energy 12460 eV,  sample-detector distance $d$ = 10 m which covers an experimental 
range in $q$ between $2.15 \times 10^{-2}$ to $0.67 \times 10^{-1}$ nm$^{-1}$. A two-dimensional 
image-intensified charge-coupled device (CCD) is employed as the detector, which allows 
the study of the (an)isotropy of the scattering. The other set of SAXS data in q range 
from 0.07 to 6 nm$^{-1}$ were obtained in our lab by a Kratky commercial 
camera. The cameras and the data treatment have been described elsewhere \cite{Din:99}.

Density measurements were done in dilute solutions using a DMA 60 apparatus supplied 
by Paar, Graz, Austria. This gave a mass density of 2.42 $\pm$ 0.12 g/cm$^3$ for the 
dissolved Laponite particles. The refractive index increment was measured using 
the Refraktometer alpha-Ref Typ 1 apparatus supplied by SLS-Systemtechnik (Freiburg, Germany) operating at the wavelength of the apparatus using for static light 
scattering (632.8 nm).

\section{Theory}
\subsection{Analysis of particle interaction by light and small-angle x-ray scattering}
We consider a dilute isotropic suspension of $N$ monodisperse platelets per volume $V$. 
The scattering intensity $I(q)$ can be rendered as the product of the intensity $I_0(q)$ 
of single platelets and the structure factor $S(q)$ through 
\cite{Guinier:55,Feigin:87,Higgins:94}
\begin{equation}
I(q) = \frac{N}{V} I_0(q) S(q)\,. 
\label{S(q)}
\end{equation}
The scattering intensity of a single platelet $I_0(q)$ may be used to define the 
form factor $P(q)$ that is normalized to unity at $q =0$ by
\begin{equation}
I_0(q) = V_p^2 ( \bar \rho - \rho_m)^2 P(q)\,,
\label{eq-P(q)}
\end{equation}
where $V_p$ denotes the volume of the particles, $\bar \rho$ is the scattering  
length density of the particles and $\rho_m$ is the respective quantity of the 
solvent (see the discussion of this point in Ref.\ \cite{Ros:02,Bal:02}). Since 
the volume fraction is given by  $\phi = (N/V)V_p$ it follows that 
$I(q=0)/\phi = V_p ( \bar \rho - \rho_m)^2$ where $\bar \rho - \rho_m$ is the contrast 
of the particles in the respective solvent. If the concentrations of the solute is 
expressed through the weight concentrations $c$, $V_p$ must be replaced by the 
molecular weight $M$, that is
\begin{equation}
I(q) = K c M P(q) S(q)\,.
\label{eq-LS}
\end{equation}
In case of SAXS the constant $K$ then follows from the number of scattering units 
per unit mass of the solute. For light scattering, $K$ is given by \cite{Tan:61}
\begin{equation}
K = \frac{4 \pi^2 \tilde{n}_0^2 (\frac{d\tilde{n}}{dc})^2}{\lambda^4 N_A}\,,
\label{eq-K}
\end{equation}
where $\tilde{n}_0$ is the refractive index of the solvent, $(d\tilde{n}/dc)$ is the 
refractive index increment, $\lambda$ denotes the wavelength of the light, and $N_A$ is 
Avogadro's number.

With complete generality, $P(q)$ may be expanded for small $q$ to give 
\cite{Guinier:55,Higgins:94}
\begin{equation}
P(q) \approx  \exp(\frac{- R_g^2q^2}{3})\,,
\label{eq-Guinier}
\end{equation}
where $R_g$ is the radius of gyration of the particles. $I_0(q)$ may hence be 
extrapolated to $q=0$. In general, a platelet can be modelled as a circular disc of 
radius $R$ and thickness $L$. For randomly oriented monodisperse platelets we have 
\cite{Guinier:55}
\begin{equation}  
P(q)=\int\limits_0^{\frac{\displaystyle{\pi}}{\displaystyle{2}}}d\vartheta\,
\sin\vartheta\left[\frac{4\sin(qL/2\cos\vartheta)\, J_1(qR\sin\vartheta)}
{q^2RL\cos\vartheta\,\sin\vartheta}\right]^2\,,
\label{eq-platelet}
\end{equation}
where $J_1$ denotes the cylindrical Bessel function of first order.

\subsection{Interaction in dilute solution}
\label{dilute}
The dependence of $S(q)$ on concentration may be expressed in the usual virial series 
by \cite{Guinier:55,Feigin:87,Higgins:94}
\begin{equation}
\frac{1}{S(q)} = 1 + 2 B_{\rm app}\phi + 3 C_{\rm app}\phi^2 +...\,,
\label{eq-virial}
\end{equation}
where the apparent virial coefficient $B_{\rm app}$ depends on $q$ through 
\cite{Guinier:55,Ros:02,Bal:02}
\begin{equation}
B_{\rm app} = 4 \frac{(\pi/6)d_{eff}^3}{V_p}(1 - \frac{1}{10}d_{eff}^2q^2 + ... )\,,
\label{eq-vircoeff}
\end{equation}
and  $d_{eff}$ is a measure for an effective diameter of interaction  in the 
dilute regime. The quantity $C_{\rm app}$ is the correction related to the 
third virial coefficient. Eq. (\ref{eq-vircoeff}) can be derived easily from 
the fact that binary interaction prevails if the concentration is low enough. 
Therefore the pair correlation function $g(r)$ follows simply by 
$g(r) = \exp[-V(r)/kT]$ \cite{hans:86}.The scattering data obtained in the 
dilute regime can therefore be extrapolated to vanishing concentration by 
plots of $\phi/I(q)$ against concentration. Moreover, Eq.~(\ref{eq-vircoeff}) 
suggests that the slope of these plots that give the apparent virial coefficient 
$B_{app}$ will diminish with increasing value of $q$ \cite{Note:1}. Hence, a 
precise analysis of the scattering intensity $I(q)$ obtained for different 
concentrations leads not only to a proper extrapolation to vanishing volume 
fraction, but also to parameters characterizing the interaction of the particles: 
Evidently, Eq. (\ref{eq-vircoeff}) refers only to equilibrium states. The 
analysis of $I(q)$ obtained in the dilute regime may therefore yield to criteria 
that allow us to exclude non-equilibrium states. 

Arguments derived from first principles demonstrate that $S(q)$ can be expanded 
into a series of powers in $q^2$ \cite{Hel:80,Apf:94,Ros:02,Wei:99}
\begin{equation}
S(q) = S(0) + \alpha q^2 +...\,,
\label{expansion}
\end{equation}
where the coefficient $\alpha$ is related to the second moment of the interaction 
potential $V(r)$ and provides a measure for the range of interaction of the 
particles in suspension (see the discussion of this point in Ref.~\cite{Wei:99}). In 
contrast to Eq.~(\ref{eq-vircoeff}), this relation holds for all possible volume 
fractions. Therefore Eq.~(\ref{expansion}) provides another check for the data. 
In particular, it is highly useful to exclude scattering intensities that have 
been afflicted by the presence of aggregates \cite{Ros:02}.

\subsection{Polydispersity}
As mentioned above, previous work has shown that Laponite particles are not 
monodisperse. The effect of polydispersity must therefore be taken into account 
and the intensity becomes an appropriate average over all species $i$
\begin{equation}
I(q) = K \frac{\sum\limits_{i=1}^s c_i M_i P_i(q)}
{\sum\limits_{i=1}^s c_i M_i} S_M(q) = K c M_w P_w(q) S_M(q)\,,
\label{Polyd}
\end{equation}
where $P_i(q)$ is the normalized form factor of a platelet of species $i$
[Eq.~(\ref{eq-platelet})], $S_M$ is the "measured structure factor", and $s$ is 
the total number of species. Moreover, we have $\sum_{i=1}^s c_i =c$. The 
weight-average molecular weight $M_w$ of the platelets follows from this as 
\begin{equation}
M_w = \frac{\sum\limits_{i=1}^s c_i M_i}{\sum\limits_{i=1}^s c_i}\,.
\label{Mw}
\end{equation}
The normalized form factor of the multicomponent system $P_w(q)$ can be expressed 
through the form factors $P_i(q)$ of platelets of species $i$ by
\begin{equation}  \label{eq6}
P_w(q)=\frac{\sum\limits_{i=1}^sc_i P_i(q)}{\sum\limits_{i=1}^s c_i}\,,
\end{equation}
again with $P_w(0)=1$. Obviously, all quantities discussed in the previous section 
as e.g. $B_{app}$ and $d_{eff}$ must be replaced by appropriate mean values.

\subsection{Multicomponent interaction site model}
A quantitative understanding of correlations and interactions 
between various colloidal species can be achieved using the well-established 
techniques of liquid-state theory \cite{hans:86}. The PRISM ("polymer reference 
interaction site model") theory, originally designed for the study of polymer 
solutions and melts \cite{schw:97}, has been recently extended to investigate 
various charged and uncharged colloidal systems 
\cite{yeth:96,shew:97,harn:00,harn:01,harn:01a,shew:02,hofm:03,zher:03}, 
and mixtures of neutral spherical colloids and polymers 
\cite{yeth:92,khal:97,chat:98,fuch:00,cost:05}, as well as like- and oppositely-charged 
colloids and polyelectrolytes \cite{ferr:00,harn:02}. The key theoretical problem 
to be addressed here is that of an interpretation of experimental 
scattering data on very dilute Laponite clay suspensions within the framework of 
PRISM. The main difference of this work from 
the previous theoretical studies on monodisperse \cite{harn:01} and bidisperse 
\cite{harn:02} platelet suspensions is that we take into account polydispersity 
in the size of the platelets by using a multicomponent PRISM integral equation 
theory. We consider a multicomponent system involving $s$ species of charged 
platelets with number densities $n_i$, where $1\le i \le s$. Each platelet 
of species $i$ contains $N_i$ equivalent interaction sites. The interaction
potential between sites on particles of species $i$ and $j$, carrying the 
charges $z_ie$ and $z_je$, will be of the generic form:

\begin{figure}[t]
\vspace*{0.3cm}
\includegraphics[width=8cm]{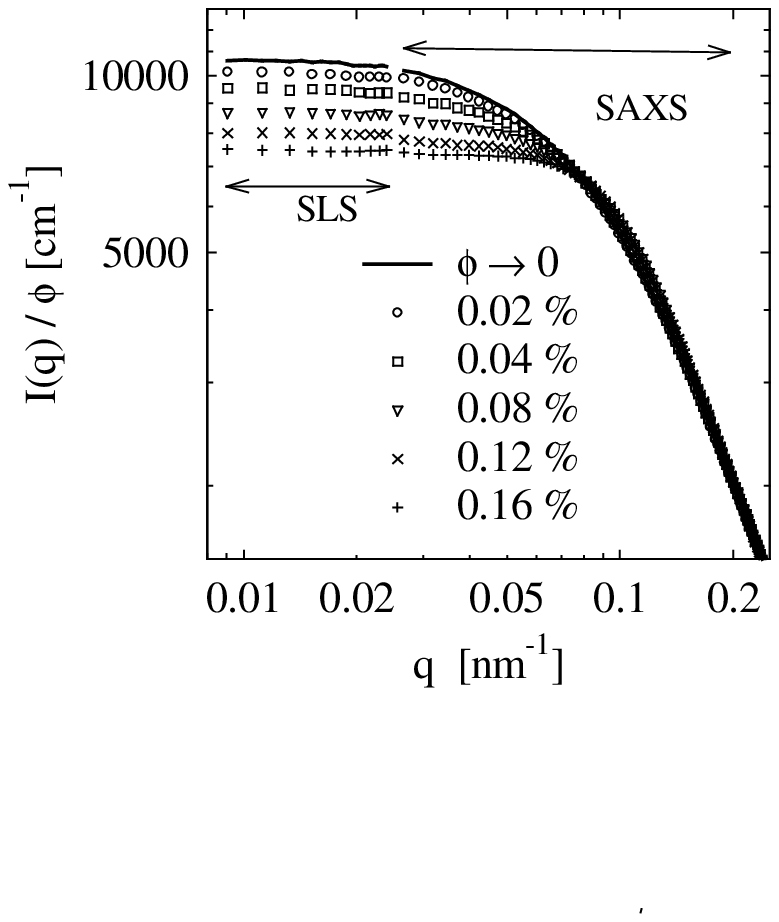}
\caption{Absolute scattering intensities of Laponite suspensions normalized to 
their volume fraction $\phi$ indicated in the graph. The data have been obtained 
by static light scattering in the region of smallest $q$-values 
($7.9 \times 10^{-3} - 2.3 \times 10^{-2}$ nm$^{-1}$). Data at higher $q$ were taken 
by using the beamline ID2 ($2.15 \times 10^{-2}$ to $0.67 \times 10^{-1}$ nm$^{-1}$) 
and by a conventional Kratky-camera ($0.07$ to $6$ nm$^{-1}$). The solid line refers 
to the intensity extrapolated to vanishing concentration.}
\label{fig1}
\end{figure}

\begin{equation}  \label{eq1}
u_{ij}(r)=\frac{\displaystyle z_iz_je^2}{\displaystyle\varepsilon r}
\exp(-\kappa_Dr)\,,
\end{equation}
where $\lambda_D=\kappa_D^{-1}$ is the usual Debye screening length.
The system under investigation is an aqueous dispersions, but in view of the 
mesoscopic scale of the platelets, the solvent is modelled as a structureless 
dielectric continuum providing a macroscopic permittivity $\varepsilon$. Any 
microscopic counterions or ions from added electrolyte will be considered at 
the linear response (or Debye-H\"uckel) level, i.e., they will screen the 
electrostatic potential due to the interaction sites on the charged platelets 
on a scale given by the Debye screening length. The underlying 
Born-Oppenheimer-like assumption entails that the charge distribution on the 
mesoscopic particles does not contribute to screening. 

PRISM is based on the assumption that all direct correlation functions between 
sites on pairs of different platelets are identical. This leaves a total of 
$s(s+1)/2$ independent direct correlation functions $c_{ij}(q)$. Corresponding 
total correlation functions $h_{ij}(q)$ are defined by averaging over the 
$N_iN_j$ correlation functions between all pairs of sites $i$ and $j$ on two 
platelets of the same or different species. The two sets of correlation 
functions are related by the generalized Ornstein-Zernike equations of 
the PRISM, which in Fourier space read \cite{schw:97}

\begin{figure}[t]
\vspace*{0.3cm}
\includegraphics[width=7.5cm]{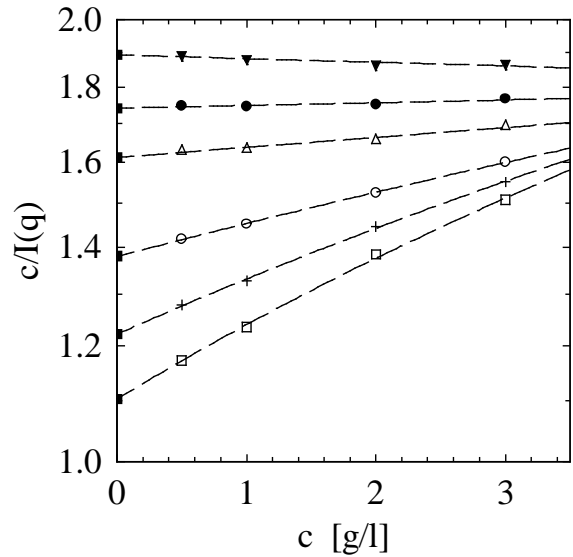}
\caption{Extrapolation of the measured scattering intensities $I(q)$ to vanishing 
concentration according to Eq.~(\ref{eq-virial}) and (\ref{eq-vircoeff}). Parameter of the curves is the magnitude $q$ of the scattering vector. The respective values are (in $nm^{-1}$): quadrangles: 0.0215; crosses: 0.037; open circles: 0.0526; open triangles: 0.0681; filled circles: 0.0759; filled trangles: 0.0837.  }
\label{fig2}
\end{figure}

\begin{equation}  \label{eq3}
h_{ij}(q)=\omega_i(q)c_{ij}(q)\omega_i(q)+
\omega_i(q)\sum_{l=1}^sc_{il}(q)n_l N_l h_{lj}(q)\,,
\end{equation}
where $N_i n_i$ is the number density of interaction sites of platelets 
of species $i$ and $\omega_i(q)=N_i P_i(q)$. The set of $s(s+1)/2$ independent
Ornstein-Zernike relations must be supplemented by as many closure relations 
between each pair of total and direct correlation functions. Here we adopt the 
Laria-Wu-Chandler closure \cite{lari:91} which has recently been used for charged 
polymers and colloids \cite{yeth:96,shew:97,harn:00,harn:01,shew:02,hofm:03}
\begin{equation}  \label{eq4}
h_{ij}(r)=\ln[h_{ij}(r)+1]+\omega_i\star[c_{ij}+(k_BT)^{-1}u_{ij}]*\omega_j\,,
\end{equation}
where the asterics $\star$ denotes a convolution product. In addition, steric
effects are taken into account using the Percus-Yevick approximation \cite{hans:86}.

The closed set of equations (\ref{eq3}) and (\ref{eq4}) are solved numerically 
by a standard iterative procedure to obtain partial structure factors 
\begin{equation}  \label{eq5}
S_{ij}(q)=n_i N_i\omega_{i}(q)+n_i n_j N_i N_j h_{ij}(q)\,.
\end{equation}
Scattering experiments measure the intensity of scattered radiation as a 
function of momentum transfer. To a multiplicative constant this quantity
is the same as the sum of the partial structure factors 
$\sum\limits_{i,j=1}^sS_{ij}(q)$ which is denoted as the static structure factor. 
Most experimental papers on polyectrolytes, however, report the ratio of the static
structure factor to the form factor. Therefore we compare in the following section
the theoretical prediction for the function
\begin{equation}  \label{eq7}
S_M(q)=\frac{\sum\limits_{i,j=1}^sS_{ij}(q)}{\sum\limits_{i=1}^s n_iN_i\omega_i(q)}
\end{equation}
to experimental data. In keeping with the experimental terminology we refer 
to $S_M(q)$ as the measured structure factor.

\begin{figure}[t]
\vspace*{0.3cm}
\includegraphics[width=8.0cm]{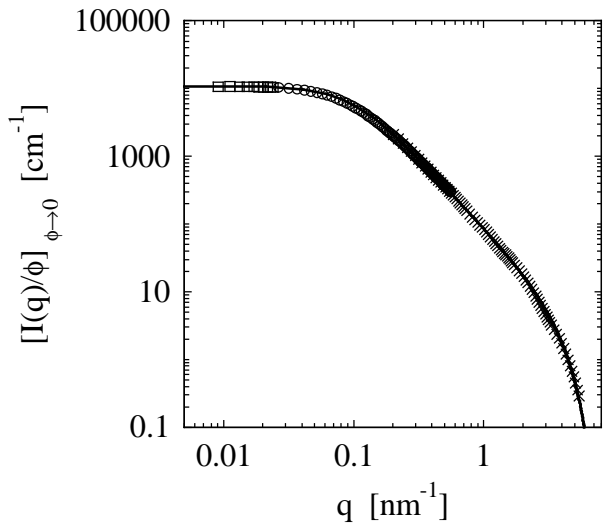}
\caption{Intensity $I_0(q)/\phi$ extrapolated to vanishing volume fraction $\phi$. 
The data have been obtained by static light scattering in the region of smallest 
$q$-values ($7.9 \times 10^{-3} - 2.3 \times 10^{-2}$ nm$^{-1}$). Data at higher $q$ were 
taken by using the beamline ID2 ($2.15 \times 10^{-2}$ to $0.67 \times 10^{-1}$ nm$^{-1}$) 
and by a conventional Kratky-camera ($0.07$ to $6$ nm$^{-1}$). The solid line marks the 
fit according to Eq.~(\ref{eq-platelet}).}
\label{fig3}
\end{figure}

\section{Results and Discussion}
\subsection{Static scattering experiments: consistency check}
As mentioned in the Introduction, the present investigation focuses on the 
dilute regime in which no gelation occurs. However, no concentration is small enough 
to disregard the influence of mutual interaction of the platelets, i.e., of $S(q)$. 
Moreover, the discussion in section \ref{dilute} has demonstrated that scattering 
data in this regime furnish valuable data as $B_{app}$ that may be used for a 
consistency check. As a first step in this analysis scattering data are therefore 
taken at small concentration ($\phi \leq 0.16$ \%) and carefully extrapolated to 
vanishing concentration. Since we aim at the limits for vanishing scattering angles 
for all quantities under consideration, the $q$-range accessible by SAXS was not 
sufficient. Hence, static light scattering was used to obtained data at smallest 
$q$-values possible in order to supplement the analysis by synchrotron SAXS. For 
the highest scattering angles the data obtained by synchrotron SAXS were supplemented 
by intensities taken by a conventional Kratky-camera. This device can take data up 
to 6 nm$^{-1}$.

\begin{figure}[t]
\vspace*{0.3cm}
\includegraphics[width=7.5cm]{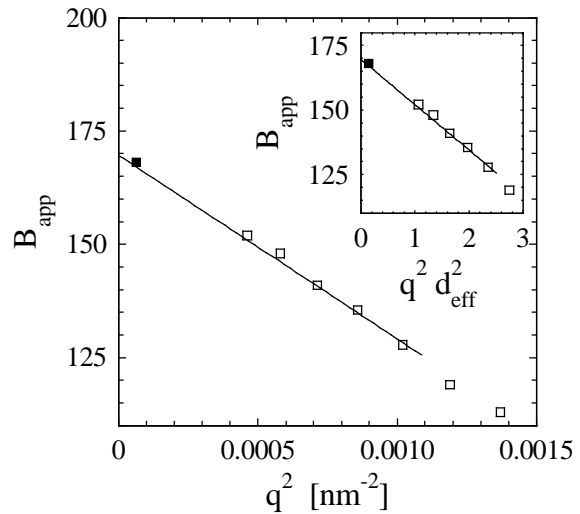}
\caption{ Apparent second virial coefficient $B_{app}$ (cf. Eq.~(\ref{eq-vircoeff})) 
of Laponite suspensions in the dilute regime. the filled circle denotes the value 
obtained by static light scattering whereas the other points have been obtained by 
SAXS. The effective diameter $d_{eff}$ deriving from this fit results to 48 nm. }
\label{fig4}
\end{figure}

In all cases absolute scattering intensities have been obtained. For light scattering, 
the constant $K$ in Eq.~(\ref{eq-K}) was found to $K =(4.99 \pm 1)~ 10^{-8}$ cm$^2$ Mol/g$^2$ 
from the measured refractive increment ($d\tilde{n}/dc = 0.083 \pm 0.008$ mL/g). In case of SAXS 
the respective optical constant $K$ in Eq.~(\ref{eq-LS}) was found to be 
$K = 0.0445$ Mol/g$^2$. All absolute intensities agreed within the present limits of 
error ($\pm 5$ \%). 

Previous investigations by Bonn et al. \cite{Bon:99b} and by Nicolai and Cocard 
\cite{Nic:00} have shown that even dilute aqueous suspensions of Laponite may contain 
aggregates that must be filtered off, in particular for light scattering experiments. Hence, all suspensions 
have been carefully filtrated prior to use. Moreover, special care was exerted in 
preparing the solutions by carefully stirring for a prolonged time. If the time for 
the process of dissolution was chosen to be short or without vigorous stirring, aggregates 
resulted that showed up in the region of smallest angles. Evidently, the pH 10 and 
the ionic strength are further parameters of central importance. Most of the data to be 
discussed here were taken from solutions in which the ionic strength was set to 
0.001 mM by NaCl.

Figure \ref{fig1} displays the data obtained at low ionic strength for various volume 
fractions $\phi$ calculated from the weight concentrations of the solute and its density 
in solution (2.42 $\pm$ 0.12 g/cm$^3$). First of all, the data obtained from SLS fit well 
together with the data obtained from SAXS. Moreover, the influence of the structure 
factor $S(q)$ cannot be disregarded even at the lowest volume fraction. This is due to 
the electrostatic stabilization that must lead to an effective diameter of interaction 
$d_{eff}$ which is larger than the largest dimension of the platelets. Figure \ref{fig2} 
demonstrates that these data can be safely extrapolated to vanishing concentration 
by use of Eq.~(\ref{eq-virial}) and (\ref{eq-vircoeff}). Plots of $c/I(q)$ vs. $c$ lead 
to straight lines. The resulting intercepts are given as solid lines in Fig.~\ref{fig1}. 
Even at the smallest concentration ($\phi=0.02$ \%) the scattering intensities
deviate from the value resulting from this extrapolation. Only at higher scattering 
angles the influence of mutual interaction is absent in full accord with 
Eq.~(\ref{eq-vircoeff}). 

\begin{figure}[t]
\vspace*{0.3cm}
\includegraphics[width=8.0cm]{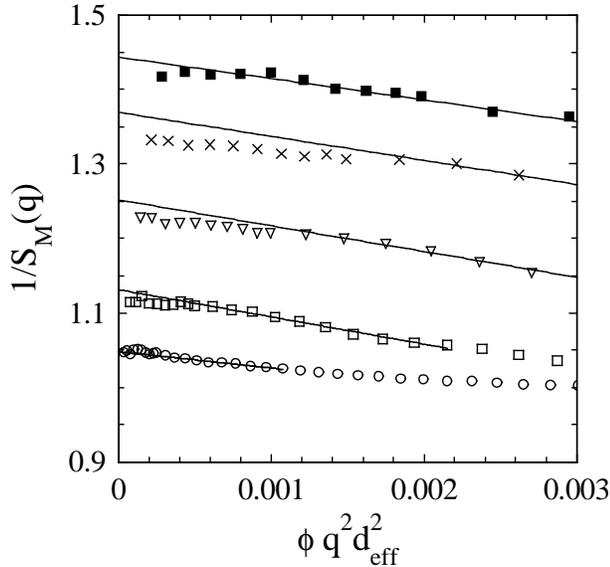}
\caption{Consistency check of the structure factor $S_M(q)$ in the region of small 
$q$-values according to Eq.~(\ref{eq-virial}) and (\ref{eq-vircoeff}). }
\label{fig5}
\end{figure}

Figure \ref{fig3} displays the intensity extrapolated to vanishing concentration for the 
entire $q$-range available from the combination of three devices, namely static light 
scattering, synchrotron radiation, and a conventional Kratky-camera. The rather flat 
region of small angles is due to the fact that $I_0(q)$ must follow Guinier's law 
(see Eq.~(\ref{eq-Guinier})) at small values of  $q$. From a Guinier plot of these data (not shown) 
we obtained a weight average of the platelets $M_w = 930 \pm 190$ kg/Mol and a radius of 
gyration $R_g = 13.4$ nm. At intermediate $q$ the slope of $-2$ is seen predicted by 
Eq.~(\ref{eq-platelet}). Finally, at highest $q$-values the scattering intensity is greatly 
diminished because of the first minimum of $P(q)$ which is located at $q =$ 7.1 nm$^{-1}$.

The form factor $P(q)$ of a thin platelet can be easily calculated according to 
Eq.~(\ref{eq-platelet}). From this fit it becomes obvious that polydispersity must be 
included to give a satisfactory fit. This is due to the fact that Eq.~(\ref{eq-platelet}) 
leads to a wavy from of $I_0(q)$ for monodisperse platelets which is smoothed out by 
polydispersity. The solid line shows the resulting fit obtained for a 
Schulz-Zimm-distribution \cite{schu:39,zimm:48} with a polydispersity expressed through 
$M_w/M_n = 1.5$. The weight-average radius $R_w$ is given by 10.5 nm, the thickness 
is found to be 0.9 nm, and the calculated weight average [Eq.~(\ref{Mw})] is in agreement 
with the measured value. We reiterate that Laponite platelets exhibit a appreciable 
polydispersity that can be deduced from this analysis. Obviously, this must be taken into 
account when modelling the interaction of the particles at finite volume fraction.

\begin{figure}[t]
\vspace*{0.3cm}
\includegraphics[width=8.0cm]{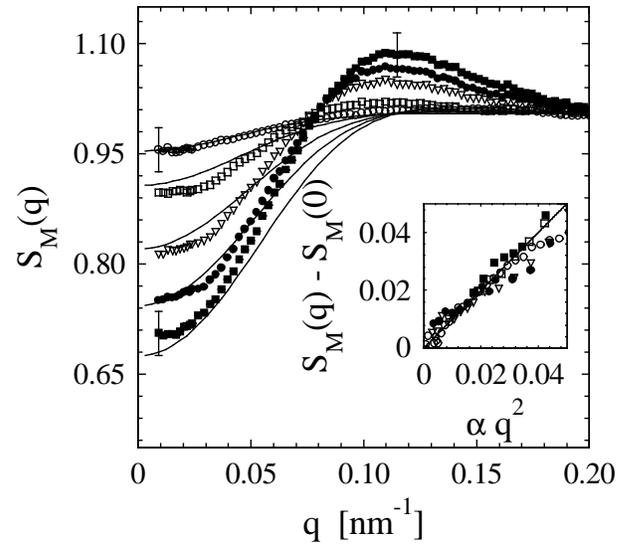}
\caption{Structure factor $S_M(q)$ determined experimentally for various 
volume fractions (open circles: 0.02 $\%$; open squares: 0.04 $\%$; 
open triangles: 0.08 $\%$; solid circles: 0.12 $\%$; solid squares: 0.16 $\%$). 
The solid lines represent the corresponding structure factors computed according 
to Eq.~(\ref{eq7}) and assuming a purely repulsive interaction between the 
platelets [Eq.~(\ref{eq1})]. The inset shows the consistency check of the 
experimental data according to Eq.~(\ref{expansion}).}
\label{fig6}
\end{figure}

The slope of the curves shown in Fig.~\ref{fig2} can now be evaluated to yield the 
apparent second virial coefficients $B_{app}$ defined through Eq.~(\ref{eq-virial}). 
Figure \ref{fig4} displays the resulting data and a fit according to Eq.~(\ref{eq-vircoeff}).
>From this fit the effective diameter $d_{eff}$ was obtained to 48 nm which is considerably 
higher than the largest dimension of the platelets (see above). This can be traced back 
to the electrostatic repulsion that is already operative before the platelets touch each 
other. Hence, $d_{eff} > 2R_w$ and $B_{app}$ cannot be calculated from the
dimensions of the platelets in solution. The inset gives the same plot but with the 
abscissa scaled by $\phi d_{eff}$ as suggested by Eq.~(\ref{eq-vircoeff}). A similar plot 
is displayed in Fig.~\ref{fig5} showing the extrapolation to vanishing concentration as 
in Fig.~\ref{fig3}. Figure \ref{fig5} demonstrates that approximately parallel lines 
result from this scaling of the abscissa. Thus, the data obtained at low concentrations 
follow Eq.~(\ref{eq-vircoeff}). This indicates that these solutions fully qualify as 
equilibrium states. 

\begin{figure}[t]
\vspace*{-1.3cm}
\includegraphics[width=9.8cm]{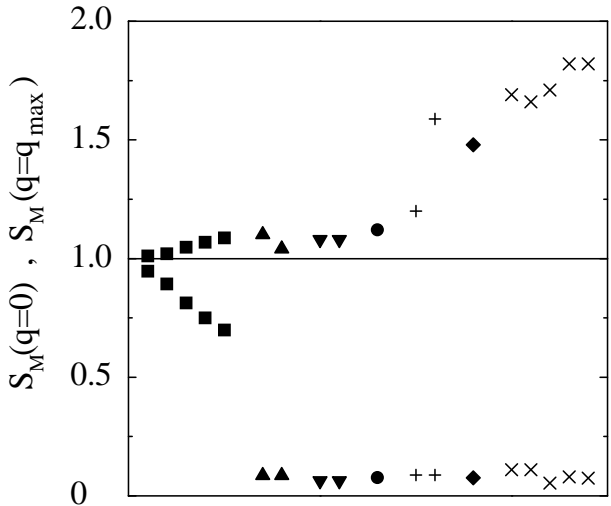}
\vspace*{-5.2cm}
\caption{The structure factor at zero scattering vector \mbox{$S_M(q=0)$} (lower symbols)
together with the structure factor at the main peak $S_M(q=q_{max})$
(upper symbols) for various suspensions: Laponite at five volume fractions: 
0.02 $\%$, 0.04 $\%$, 0.08 $\%$, 0.12 $\%$, 0.16 $\%$ from left to right (squares);
polystyrenesulfonate of length $L=6.8$ nm at two concentrations: 0.1 mol/L and
0.2 mol/L from left to right (up triangles \cite{kass:97}); 
polystyrenesulfonate of length $L=40$ nm at two concentrations: 0.1 mol/L and
0.2 mol/L from left to right (down triangles \cite{kass:97});
DNA of length $L=57$ nm at $0.05$ mol/L (circles \cite{kass:97});
DNA of length $L=380$ nm at $0.05$ mol/L (diamonds \cite{kass:97});
proteoglycan at two different salt concentrations: 0.05 mM and 0 mM 
from left to right (plus-symbols \cite{norw:96});
tobacco mosaic virus at five concentrations: 
0.11 mg/mL, 0.27 mg/mL, 0.43 mg/mL, 1.05 mg/mL, 2.07 mg/mL
from left to right (crosses \protect\cite{maie:88}).}
\label{fig8}
\end{figure}

The intensity referring to $\phi \longrightarrow 0$ may now be used to determine the 
structure factor $S(q)$ for nonzero values of $\phi$. The foregoing discussion has 
clearly revealed, however, that the Laponite particles exhibit an appreciable 
polydispersity. The structure factor obtained from this division should therefore be 
treated as a "measured" quantity $S_M(q)$ that contains explicitly a contribution of 
the size distribution. Figure \ref{fig6} displays the resulting $S_M(q)$ obtained for 
the low ionic strength of $0.001$ mM. The inset shows the consistency check according to 
Eq.~(\ref{expansion}). It underscores again the fact that we are dealing with equilibrium 
states. Figure \ref{fig6} shows that $S_M(q)$ is considerably lowered at vanishing scattering
angle and exhibit a shallow maximum around $q = 0.11$ nm$^{-1}$. As mentioned above, most 
of the investigations using scattering methods have been conducted near the concentration
of gelation, i.e., at much higher concentrations, and the structure factor has not been 
extracted. Hence, we cannot compare the present data directly with similar measurements 
from literature. 

\begin{figure}[t]
\vspace*{0.3cm}
\includegraphics[width=8.3cm]{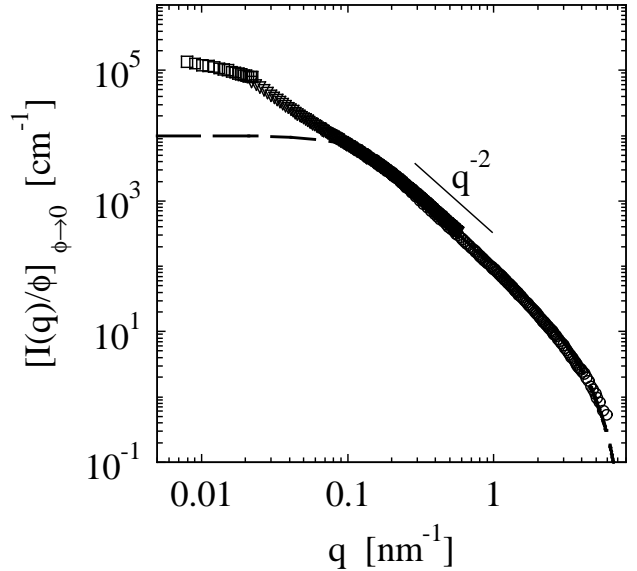}
\caption{Slow aggregation of Laponite particles at higher ionic strength (0.005 mM). 
The intensity $I_0(q)/\phi$ was extrapolated to vanishing volume fraction $\phi$ 
(cf. the discussion of Fig.~\ref{fig3}). Note the much higher intensity obtained at 
small scattering vectors as compared to the case of the lower ionic strength (0.001 mM)
in Fig.~\ref{fig3}.}
\label{fig7}
\end{figure}

The results for the structure factor as calculated according to 
Eqs.~(\ref{eq1}) - (\ref{eq7}) are compared to the experimental scattering data in 
Fig.~\ref{fig6}. The adjustable parameter related to the potential of interaction 
is the effective charge of the platelets which is much smaller than the structural 
charge of the platelets, due to counterion "condensation" 
\cite{delv:99,triz:02,levi:98}. It is the strong reduction of the bare charge 
which allows the use of linearized Poisson-Boltzmann theory, leading to the 
simple exponential Debye screening of the Coulomb interaction [Eq.~(\ref{eq1})]. 
In the calculations we use a fixed surface charge density which is defined as the 
charge of a platelet of species $i$ divided by the area $A_i=\pi R_i^2$ of the face 
of the platelet. The calculated structure factors agree with the experimental data 
for small scattering vectors, while pronounced deviations are visible in the intermediate 
scattering vector regime. Particularly, the theoretical calculations predict 
considerably smaller peaks of the structure factor than observed experimentally. 
Increasing the surface charge density, as compared to the one used in Fig.~\ref{fig6},
leads to an increase of the height of the peaks of the calculated structure factors 
and to a decrease of the structure factors at small scattering vectors $q\to 0$. 
We emphasize that neither using different model parameters (surface charge density, 
Debye screening length, and size of the platelets) nor using a different distribution 
function (Gaussian, log-normal, and Tung distribution) lead to an agreement with the 
experimental data. 

Since structure factors and pair correlation functions of polyelectrolyte solutions
calculated within the same theoretical framework are in good agreement with experimental 
data and computer simulations \cite{yeth:96,shew:97,harn:00,harn:02,shew:02}, it is 
worthwhile to compare scattering data for polyelectrolytes with the data obtained 
for Laponite. Figure \ref{fig8} displays $S_M(q=0)$ which is proportional to the isothermal 
compressibility together with $S_M(q=q_{max})$ for various polyelectrolytes and Laponite, 
where $q_{max}$ is the absolute value of the scattering vector at the main peak. The 
structure factors of the polyelectrolytes exhibit a main peak $S_M(q=q_{max})>1$ and 
a {\it small} value $S_M(q=0)<0.11$ for small scattering vectors, while the structure 
factors of Laponite are charcterized by a main peak and a rather {\it large} value 
$0.7<S_M(q=0)<0.95$ for small scattering vectors. The enhancement of the small angle 
($q\to 0$) value of $S_M(q)$ of Laponite as compared to the polyelectrolytes signals 
increased density 
fluctuations. Since the polyelectrolyte solutions remain liquid-like even at higher 
concentrations, the observed qualitative different behavior of the structure factors of 
Laponite may be considered as indicative of the sol-gel transition at higher concentrations. 
Notice that the values of $S_M(q=0)$ for the Laponite suspensions under considerations are 
{\it smaller} than 1 and {\it decrease} with increasing concentration in contrast to 
strong small $q$ upturns which have been observed experimentally for clay suspensions 
in the gel phase \cite{morv:94,pign:97,Mou:98b,Kro:98,saun:99,Bon:99a,levi:00,bhat:03}, 
polyelectrolyte gels \cite{scho:94} and mixtures \cite{norw:96}, 
polystyrenesulfonate ion exchange resins \cite{maar:96}, and  low ionic 
strength polyelectrolyte solutions \cite{foer:90,boue:94,ermi:98}. These strong
upturns signal strong concentration fluctuations indicative of aggregation, 
or spinodal instability reminiscent of the behaviour observed in recent computer 
simulations \cite{lins:99} and PRSIM integral equations \cite{harn:02} of mixtures 
of oppositely charged  particles. 

A final point to be discussed at this point is the stability of Laponite suspensions
at higher ionic strength. Figure \ref{fig7} displays the intensity obtained for an ionic 
strength of $0.005$ mM. The same procedures have been applied as in case of the data 
discussed above. The intensities have been extrapolated to vanishing concentration in 
order to remove all alterations due to $S_M(q)$. The dashed line marks the form factor 
obtained previously for $0.001$ mM as described above. It is obvious that the intensity 
measured at small scattering angles is considerably higher than in case of the lower 
ionic strength. We interpret this as the onset of a slow coagulation process. Hence, a 
salt concentration as low as $0.005$ mM already suffices to lead to the built-up of 
aggregates visible in the scattering experiment at small $q$. This points clearly to the 
limited stability of the Laponite platelets in solutions which will be further discussed 
below.

\subsection{Effective interaction potential}
The results of the preceding subsection indicate that the disagreement between the
experimental and theoretical results may be due the purely repulsive screened Coulomb
potential used in the calculations [Eq.~(\ref{eq1})]. Recently Ruzicka et al. 
\cite{ruzi:04} have suggested that at low concentrations the effective interaction 
between Laponite platelets are characterized by a competition of long-range 
Coulomb repulsion and short-range  van der Waals attraction similar to colloidal 
systems and protein solutions \cite{piaz:00,pell:03,scio:04}. Moreover, the charge 
density is higher on the face than on the rim of a platelet, which leads to a 
modification of the interaction potential between sites on particles $i$ and $j$ 
[Eq.~(\ref{eq1})] within the computationally very demanding multicomponent PRISM theory. 
Here we attempt a mesoscopic coarse graining, whereby particles act via an effective 
potential $V(r)$. To this end we solve numerically the Ornstein-Zernike equation

\begin{figure}[t]
\vspace*{-1.3cm}
\includegraphics[width=9.8cm]{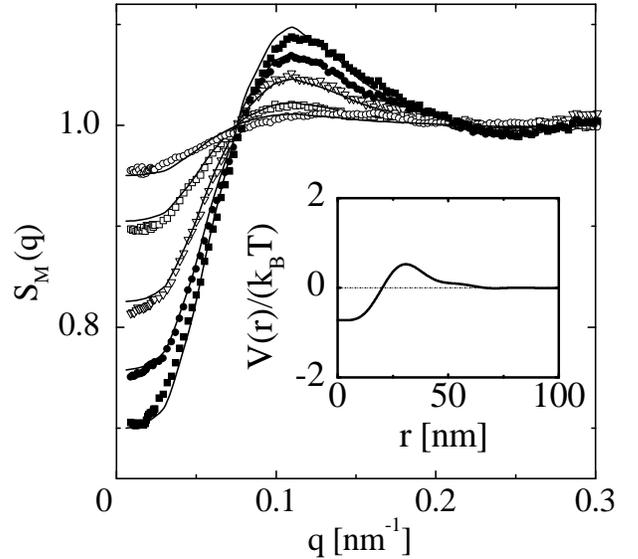}
\vspace*{-5.2cm}
\caption{Structure factor $S_M(q)$ determined experimentally for various 
volume fractions (open circles: 0.02 $\%$; open squares: 0.04 $\%$; 
open triangles: 0.08 $\%$; solid circles: 0.12 $\%$; solid squares: 0.16 $\%$). 
The solid lines represent the results of the integral equation theory 
[Eqs.~(\ref{eq20}) - (\ref{eq22})] using the effective potential $V(r)$ shown in 
the inset.}
\label{fig9}
\end{figure}

\begin{equation} \label{eq20}
h(q)=c(q)+nc(q)h(q)
\end{equation}
together with hypernetted-chain closure relation
\begin{equation} \label{eq21}
h(r)=\ln[h(r)+1]+c(r)+(k_BT)^{-1}V(r)\,.
\end{equation}
These equations follow from Eqs.~(\ref{eq3}) and (\ref{eq4}) for $s=1$, $\omega_i(q)=1$, 
and $N_i=1$. The calculated structure factor 
\begin{equation} \label{eq22}
S(q)=1+nh(q)
\end{equation}
is compared in Fig.~(\ref{fig9}) with the experimental data. For all volume 
fraction, good agreement with experiment is achieved with a density-independent
effective potential $V(r)$ that is attractive for small distances but repulsive 
for larger distances (see the inset in Fig.~(\ref{fig9})). We emphasize that the 
effective potential fulfills the stability condition $\int_0^\infty dr\,r^2V(r)>0$
discussed by Ruelle \cite{ruel:69}, and provides a representation of the underlying 
many-body interactions in the system. Thermodynamic properties can be calculated 
from the compressibility relation $S(q=0)=nk_BT \kappa_T$, where $\kappa_T$ is 
the isothermal compressibility \cite{hans:86}. The osmotic pressure is 
evaluated as $P=k_BT\int dn\, S(q=0)^{-1}$.
Moreover, the correlation functions obtained from the Ornstein-Zernike equation can be 
used as input for a Langevin equation for the dynamic structure factor \cite{harn:01a} 
allowing one to study dynamical properties.

Finally, it is instructive to rationalize our results using simple arguments. The 
repulsion due to the overlap of the counterion diffusive layers can be replaced by a
renormalized hard-core repulsion with an effective diameter equal to 
$2(R_w+\lambda_D)=41$ nm, where the Debye screening length has been approximated
by $\lambda_D=(8\pi l_B ns)^{-1/2}$, where $l_B=0.714$ nm is the Bjerrum length and 
$2n_s$ is the particle number density of salt ions at ionic strength 0.001 mM. This 
effective diameter is somewhat smaller than $d_{eff}=48$ nm obtained from 
Eq.~(\ref{eq-vircoeff}). Another length of interest is the estimated average 
distance between two platelets defined as $d_h=(\phi/V_w)^{-1/3}$, where 
$V_w=\pi R^2_w L$ is the average particle volume. For \mbox{$\phi=0.02$ \%} we find 
$d_h=116$ nm and for $\phi=0.16$ \%,  $d_h=58$ nm. As is apparent from 
Fig.~\ref{fig9}, the correlation length associated with the main peak of the 
structure factor, $d_c=2\pi/q_{max}\approx 53$ nm is {\it independent} of the volume fraction 
in contrast to results for Laponite at very low ionic strength (0.0001 M; \cite{levi:00}). 
The correlation length $d_c$ for the samples shown in Fig.~\ref{fig9} is related to the 
effective diameter $d_{eff}$ rather than to average distance between two platelets 
defined as $d_h$. However, the fact that $d_{eff}$ is comparable to $d_h$ for 
$\phi=0.16$ \% indicates that the sol-gel transition occurs already at a volume fraction
slightly larger than \mbox{$\phi=0.16$ \%} for which $d_{eff}\approx d_h$, in agreement 
with our experimental findings. The main peak at the scattering vector $q_{max}$ for
gel systems reflects the short interparticle distance between connected particles of 
effective diameter. The pronounced excess scattering at small values of $q$ is due to 
correlations associated with the network structure.

\section{Conclusion}
A precise analysis of the static scattering intensity of dilute aqueous Laponite 
solutions has been presented. The combination of static light scattering with small-angle 
x-ray scattering leads to a wide range of the magnitude $q$ of the scattering vector so 
that reliable data of the structure factor $S_M(q)$ could be obtained. Special care was 
taken to show that the present data refer to true equilibrium states. The form factor
obtained from the scattering intensity extrapolated to vanishing volume fraction agrees 
with the calculated one for polydisperse platelets (see Figure \ref{fig3}). The structure 
factor $S_M(q)$ is characterized by a main Coulomb peak \mbox{$S_M(q=q_{max})>1$} and a value 
$0.7<S_M(q=0)<0.95$ for small scattering vectors corresponding to a rather large isothermal 
compressibility, in contrast to various polyelectolyte solutions (see Figure \ref{fig8}). 
The analysis of $S_M(q)$ which includes the marked effect of polydispersity demonstrates 
unambiguously that the Laponite platelets interact via a potential that is attractive on 
short distances but repulsive on longer distances (see inset of Figure \ref{fig9}). This 
clearly points to the role of short-range attraction for the process of coagulation at 
higher volume fractions. Moreover, it explains that higher concentration of added salt 
lead to aggregation because of the increased screening of the repulsive Coulomb interaction 
between the Laponite platelets (see Figure \ref{fig7}).

\vspace*{1.0cm}

\section {Acknowledgment}

\vspace*{-0.8cm}
The authors gratefully acknowledge financial support by the Deutsche 
Forschungsgemeinschaft, 
Forschergruppe "Peloide". We acknowledge allocation of beamtime 
by the European Synchrotron Radiation Facility, Grenoble, France.


\begin{thebibliography}{99}
\bibitem{Note:1} This is in contrast to the well-known Zimm-theory of dilute polymer 
solutions (see Ref. \cite{Higgins:94} and further citations given there).  
In this theory the polymer chain is assumed to be infinitely thin, that is, 
$D \approx 0$. As a consequence of this, $B_{app}$ coincides with its 
thermodynamic limit for all $q$-values.
\end{thebibliography}
\end{document}